\documentclass[iop]{emulateapj}
\usepackage{natbib,graphicx,amsmath,xspace,hyperref}

\newcommand{\pg}{PG~1211+143\xspace}
\newcommand{\nustar}{{\it NuSTAR}\xspace}
\newcommand{\kms}{\mbox{km s$^{-1}$}\xspace}

\shorttitle{A \nustar Observation of \pg}
\shortauthors{Zoghbi et al.}

\begin{document}

\title{\nustar Reveals Relativistic Reflection but No Ultra-Fast Outflow in the Quasar \pg}

\author{A.~Zoghbi \altaffilmark{1}, J.~M.~Miller\altaffilmark{1}, D. J. Walton\altaffilmark{2,3}, F. A. Harrison\altaffilmark{3}, A. C. Fabian\altaffilmark{4}, C. S. Reynolds\altaffilmark{5}, S. E. Boggs\altaffilmark{6}, F. E. Christensen\altaffilmark{7}, W. Craig\altaffilmark{6}, C. J. Hailey\altaffilmark{8},  D. Stern\altaffilmark{2}, W. W. Zhang\altaffilmark{9}}.

\altaffiltext{1}{Department of Astronomy, University of Michigan, 1085 South University Avenue, Ann Arbor, MI 48109, USA}
\altaffiltext{2}{Jet Propulsion Laboratory, California Institute of Technology, Pasadena, CA 91109, USA}
\altaffiltext{3}{Space Radiation Laboratory, California Institute of Technology, Pasadena, CA 91125, USA}
\altaffiltext{4}{Institute of Astronomy, University of Cambridge, Madingley Road, Cambridge CB3 OHA, UK}
\altaffiltext{5}{Department of Astronomy, University of Maryland, College Park, MD 20742-2421, USA}
\altaffiltext{6}{Space Science Laboratory, University of California, Berkeley, California 94720, USA}
\altaffiltext{7}{DTU Space. National Space Institute, Technical University of Denmark, Elektrovej 327, 2800 Lyngby, Denmark}
\altaffiltext{8}{Columbia Astrophysics Laboratory, Columbia University, New York, New York 10027, USA}
\altaffiltext{9}{NASA Goddard Space Flight Center, Greenbelt, Maryland 20771, USA}


\email{abzoghbi@umich.edu}

\begin{abstract}
We report on four epochs of observations of the quasar \pg
using \nustar.  The net exposure time is 300~ks.  Prior work on this source found suggestive
evidence of an ``ultra-fast outflow'' (or, UFO) in the Fe K band, with
a velocity of approximately $0.1c$.  The putative flow would carry
away a high mass flux and kinetic power, with broad implications for
feedback and black hole-galaxy co-evolution. \nustar detects \pg out to 30~keV, meaning that the continuum is well-defined
both through and above the Fe K band.  A characteristic relativistic
disk reflection spectrum is clearly revealed, via a broad Fe K
emission line and Compton back-scattering curvature.  The data offer
only weak constraints on the spin of the black hole. A careful search for UFO's show no significant absorption feature above $90\%$ confidence. The limits are particularly tight when relativistic reflection is included. We discuss the statistics and the implications of these results in terms of connections between accretion onto quasars, Seyferts, and stellar-mass black holes, and feedback into their host environments.
\end{abstract}

\section{Introduction}
The observation of a relation between the masses of supermassive black holes at the centers of galaxies and the stellar velocity dispersion \citep[$M-\sigma$ relation; ][]{2000ApJ...539L...9F,2009ApJ...698..198G} suggests a direct link between black holes and their host galaxies. Energy and momentum driven out from the central regions push gas and dust away, halting star formation and stopping AGN fueling (\citealt{1998A&A...331L...1S,2005Natur.433..604D,2005MNRAS.363L..91C}; see \citealt{2012ARA&A..50..455F} for a review). The action of AGN feedback could be achieved through the powerful radio jets in the kinetic mode \cite[e.g.][]{2007ARA&A..45..117M}. In the radiative mode, accretion disks drive powerful winds that could contribute significantly to the energy budget of the BH-galaxy system. 

Observing the properties of such a wind is of great importance, particularly in X-rays where most of the radiation from the expelled material is produced. Although warm absorber winds are common in the X-ray spectra of AGN \citep{1997MNRAS.286..513R,2003ARA&A..41..117C,2005A&A...431..111B}, with outflow velocities of $\sim 1000$ \kms and column densities of $\log(N_{\rm h})\sim20-23$ cm$^{-2}$, they are weak, providing only $\sim 0.01\%$ of the AGN bolometric luminosity \citep{2005A&A...431..111B}. The more powerful winds seen in several objects with outflow velocities of $\sim0.1c$ and column densities of $\log(N_{\rm h})=24$ cm$^{-2}$ could carry power that is a few percent of the bolometric luminosity \citep{2003MNRAS.345..705P,2009MNRAS.397..249P,2003ApJ...593L..65R,2009A&A...504..401C,2009ApJ...706..644C,2010A&A...521A..57T,2012MNRAS.422L...1T,2013MNRAS.430...60G}. 

These ultra-fast outflows (UFO) seem to be present in at least $35\%$ of observed AGN in X-rays \citep{2010A&A...521A..57T}. However, this number could be an overestimate when alternative modeling and more conservative statistical analyses are considered \citep{2006ApJ...636..674K,2008MNRAS.390..421V}. Establishing how common these outflows are, their physical and geometrical properties is therefore crucial to understanding their contribution to the energy and momentum budget of black holes and their hosts. In this letter, we present analysis of the \nustar \citep{2013ApJ...770..103H} observation of the quasar \pg. \nustar band ($3-79$ keV) with the unprecedented sensitivity at hard ($>10$ keV) X-rays, fixes the continuum and thus allows a meaningful search for blue-shifted absorption below 10 keV.

\begin{figure*}
\centering
 \includegraphics[height=170pt,clip ]{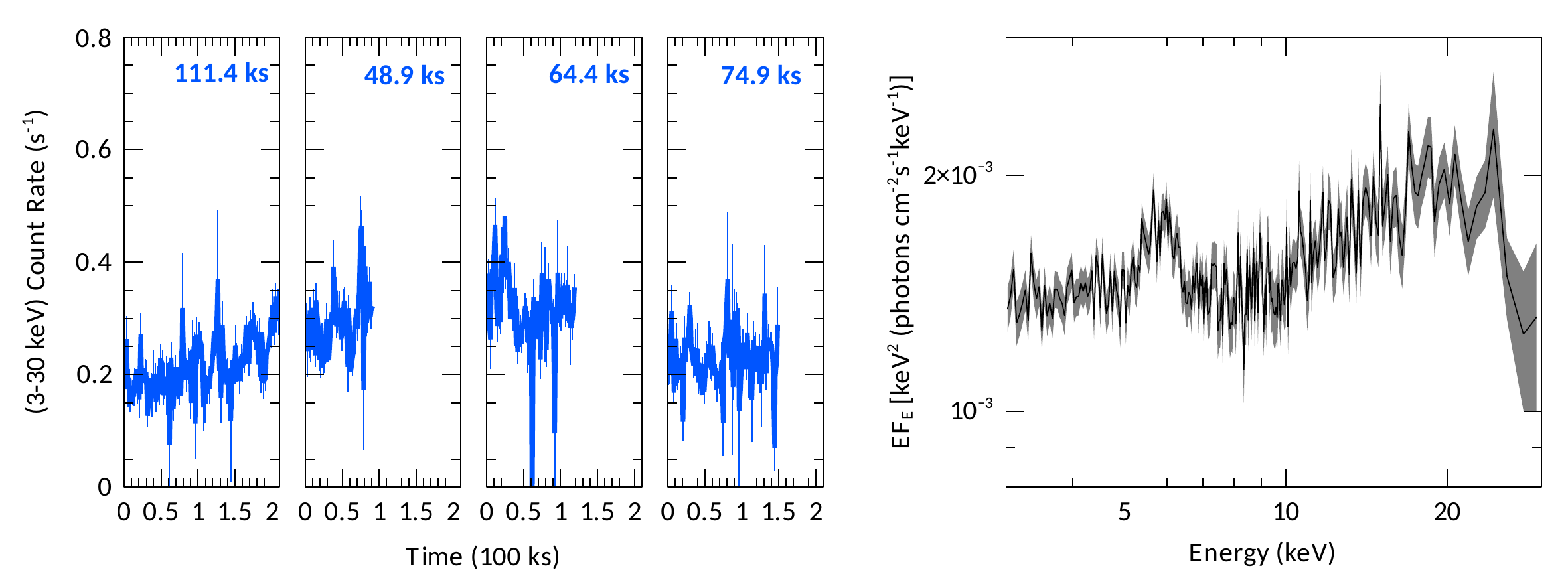} 
\caption{{\it Left:} 3--30 keV light curves from all the four \nustar observations of \pg. The net exposure is shown in each panel. {\bf Right:} The spectrum from the combined observations (combining FPMA and FPMB from all observations) unfolded to a constant to remove the effective area of the detector, and plotted as $EF(E)$ using \texttt{plot eeufspec} in \textsc{xspec}.}

\label{fig:lc_spec}
\end{figure*}

\pg ($z=0.0809$) is the archetypical case for the ultra-fast outflows in active galaxies. The first observation with {\it XMM-Newton} in 2001 showed evidence for highly blue-shifted absorption lines that are reminiscent of mildly relativistic disk winds \citep[$\sim0.1c$;][]{2003MNRAS.345..705P}. The same dataset was analyzed by \cite{2006ApJ...636..674K} who find a best fit outflow velocity of 3000 \kms instead of the high 24,000 \kms. A {\it Chandra} LETG observation showed two redshifted (instead of blueshifted!) absorption lines at 4.56 and 5.33 keV in the source frame \citep{2005ApJ...633L..81R}, which, when identified as the H-like K$\alpha$ lines corresponds to inflowing velocities of $(0.2-0.4)c$. Later {\it XMM-Newton} observations in 2004 and 2007 showed weaker lines but seem to be consistent with the original observations \citep{2009MNRAS.397..249P}, or possibly with no absorption lines at all \citep{2010A&A...521A..57T}

\section{Observations \& Data Reduction}
\nustar observed \pg in four exposures between February and July 2014 (The exact dates are: 18 February 2014, 08 and 09 April 2014 and 07 July 2014). The four observations had net exposures 111, 48, 64 and 74 ks, totaling to nearly 300 ks. The data were reduced using \textsc{heasoft v6.16} with the latest calibration (version 20141020). We used the scripts \texttt{nupipeline} and \texttt{nuproducts} to extract the spectral products. Source and background spectra were extracted from regions on (with a radius of 2 arcmin) and off source respectively and grouped so that there are least 100 source counts per bin. The spectra were analyzed using \textsc{xspec v12.8.2}. Spectral analysis was performed on individual and combined spectra as discussed in section \ref{sec:spec_analysis}. Spectra from the two focal point modules A and B (FPMA and FPMB) and from different epochs were combined using \texttt{addspec} tool in \textsc{heasoft}. The response files were combined using \texttt{addrmf} with the proper weighting.

The resulting 3--30 keV light curves from the four exposures are shown in Fig. \ref{fig:lc_spec}-left. The average 3--10 flux (from a power-law fit) is $2.7\times10^{-12}$ ergs cm$^{-2}$ s$^{-1}$ which is about the same as the first {\it XMM-Newton} observation of 2001, which had a 3--10 keV flux of $2.5\times10^{-12}$ ergs cm$^{-2}$ s$^{-1}$. The source showed some flux variability between observations. No strong spectral changes are seen apart from a normalization change in the main power-law continuum (see section \ref{sec:spec_analysis}).

\section{Spectral Analysis}\label{sec:spec_analysis}
One of the goals of the \nustar observation was to search for absorption lines from high velocity outflows. The spectrum from the new datasets is shown in Fig. \ref{fig:lc_spec}-right. It has a clear iron K emission line and an excess above 10 keV that is most likely due to the Compton reflection hump. To be systematic in the search, we consider several baseline models, including simple fits to the 3-10 keV band so we can directly compare with the baseline model used in \cite{2010A&A...521A..57T}.

\subsection{Searching for Emission and Absorption Features}\label{sec:line_search}
In the following discussions, we search for absorption (and emission) features by adding a narrow\footnote{If a broad component is present in the residuals, several neighboring narrow lines will provide a $\chi^2$ improvements.} Gaussian line and doing a systematic scan of the residuals for each baseline model. We use a grid of energies between 3.5 and 9.5 keV in 100 eV steps, and normalization values between $-1\times10^{-5}$ and $1\times10^{5}$ (to account for both emission and absorption).  We use a Monte Carlo method to obtain significance estimates. As pointed out in \cite{2002ApJ...571..545P}, a simple {\it F}-test done by comparing the improvement in $\chi^2$ after adding a Gaussian line is {\it not} appropriate for obtaining meaningful statistical confidence intervals. The baseline model is not known a priori, and neither is the reference (null) distribution of the $F$-statistic. 

\begin{figure*}
\centering
 \includegraphics[height=180pt,clip ]{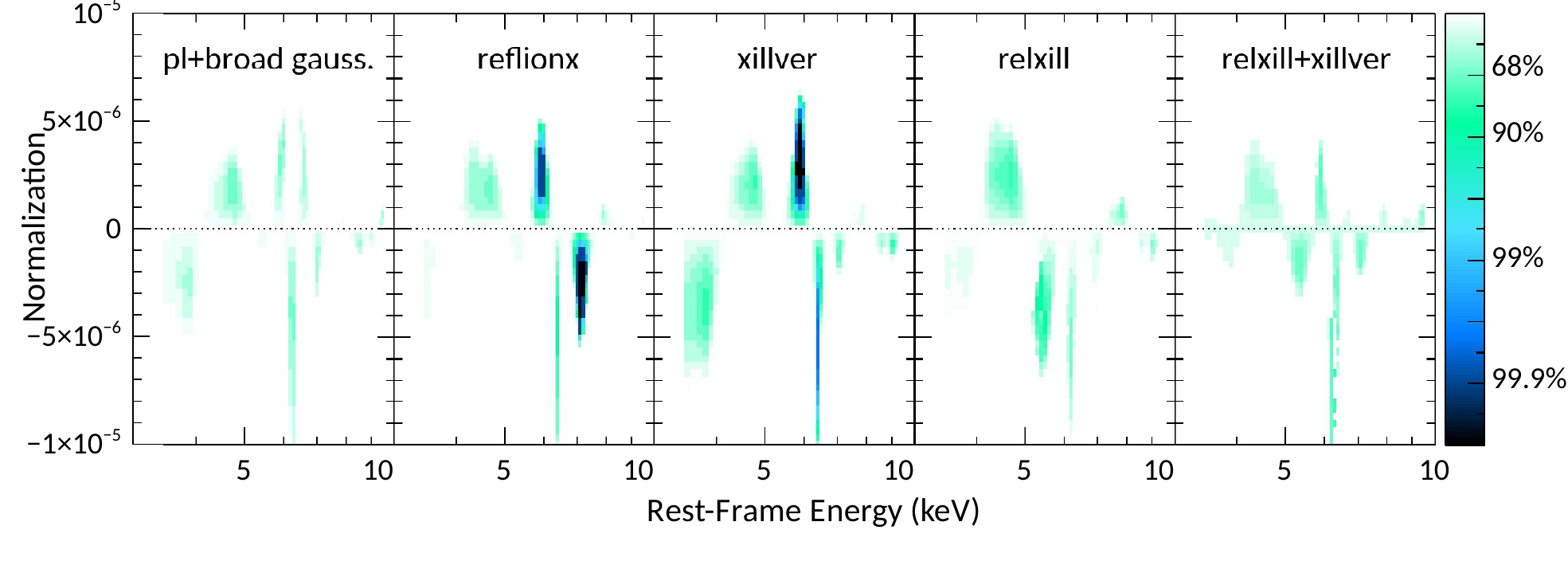} 
\caption{A line search significance plot for five baseline models. The regions with darkest colors indicate that the null model (baseline without an additional emission or absorption feature) can be rejected with that significance. The color scale corresponds to $log(1-significance)$ and the corresponding significance values are shown in the color bar. The energy axis corresponds to the source reset frame energy ($z=0.0809$).}

\label{fig:pg1211_sig}
\end{figure*}

This inapplicability of the {\it F}-test is not related to the quality of data, but rather it is fundamental to any case of an added spectral component. Using Monte Carlo methods provides a way of {\it estimating} the unknown reference distribution and then using it to assess the significance of deviations from the null model. The following steps are used to obtain a reference distribution that takes into account both the uncertainty in the baseline model parameters. These are similar to those in \cite{2004A&A...427..101P} and \cite{2010A&A...521A..57T} but additionally take into account the baseline model uncertainties:\\
A) Start with a \texttt{baseline} model with its best fit $\chi^2_0$ and covariance matrix that encodes the uncertainty of best value parameters.\\
B) Draw $N$ random sets of parameters using the covariance matrix ($N=1000$ in our case). This can be achieved for instance by running a Markov Chain Monte Carlo (MCMC) parameter search and taking parameters directly from the resulting chains (after they have converged).\\
C) For each parameter set, fake spectra using the observed response and background files. This produces $N$ simulated spectra that are drawn from the baseline model taking into account its uncertainty.\\
D) Add a narrow Gaussian line and scan in energy and normalization and record the maximum improvement in $\Delta\chi^2$. This gives $N$ values of $\Delta\chi^2$ that are used to construct a reference distribution to which observed $\Delta\chi^2$ values are compared. The significance of an observed $\Delta\chi^2_{obs}$ is $1-N_{(\Delta\chi^2>\Delta\chi^2_{obs})}/N$, where $N_{(\Delta\chi^2>\Delta\chi^2_{obs})}$ is the number of simulated $\Delta\chi^2$ values greater than $\Delta\chi^2_{obs}$.

Several baseline models are used to assess their effect on estimating absorption features:
\begin{enumerate}
\item\texttt{powerlaw + zgauss}. This consists of a power-law plus a broad Gaussian fitted to the 3--10 keV spectrum similar to those used in \cite{2010A&A...521A..57T} and \cite{2013MNRAS.430...60G}.
\item\texttt{powerlaw+reflionx} \citep{2005MNRAS.358..211R}.
\item\texttt{powerlaw+xillver} \citep{2014ApJ...782...76G}.
\item\texttt{relxill} \citep{2013MNRAS.430.1694D}.

\end{enumerate}
For models in 2,3 and 4, the whole \nustar band where the source is significantly detected above background is used. In all models, Galactic absorption is fixed at $2.7\times10^{20}$ cm$^{-2}$. For the relativistic reflection models, we assume a power-law emissivity and we fit for its index ($q$) along with the black hole spin, photon index, ionization parameter, iron abundance, inclination, high energy cutoff and reflection fraction. The details of the model parameters are shown in Table \ref{tab:params}. For each baseline model, we repeat steps A-D. The results are shown in Fig. \ref{fig:pg1211_sig}.

\begin{table}
\centering
\begin{tabular}{|l|l|}
\hline
& \texttt{powerlaw+zgauss} (3--10 keV) \\
$E_{\rm zgauss}$ & $6.28\pm0.07$ keV \\
$\sigma_{\rm zgauss}$ & $0.33\pm0.07$ keV \\
\hline
& \texttt{powerlaw+reflionx} (3--50 keV)\\
$A_{\rm Fe}$ (solar) & $0.7\pm0.1$ \\
$\Gamma$ & $2.45\pm0.06$ \\
$log(\xi)$ & $1.9\pm0.2$\\
\hline
&\texttt{powerlaw+xillver} (3--50 keV)\\
$A_{\rm Fe}$ (solar) & $0.5\pm0.1$ \\
$\Gamma$ & $2.50\pm0.06$ \\
$\log(\xi)$ & $1.3\pm0.4$\\
\hline
&\texttt{relxill} (3--50 keV) \\
$q$ & $2.2\pm.5$ \\
$a$ & $-0.13\pm0.8$ \\
$incl.$ & $28\pm7$ \\
$\Gamma$ & $2.51\pm0.2$ \\
$log(\xi)$ & $1.3\pm0.9$ \\
$A_{\rm Fe}$ (solar) & $0.7\pm0.1$ \\
$E_{\rm cut}$ & 124 (keV) (lower limit)\\
Ref. fract. & $2.9^{+1.6}_{-0.4}$ \\
\hline

\end{tabular}
\caption{Fit parameters for the baseline models discussed in section \ref{sec:line_search}.}
\label{tab:params}
\end{table}

\subsection{Line Search Results}
It is clear from Fig. \ref {fig:pg1211_sig} that there is {\it no} significant absorption feature that is persistent with all baseline models. The phenomenological description of the first model gives an excellent description to the data in the 3 -- 10 keV band with no significant residuals. In the \texttt{reflionx} fit, an absorption feature at 7.1 keV is apparent along with the emission feature at 5.8 keV. The same feature at 5.8 keV persists for the \texttt{xillver} fit.  Both the \texttt{reflionx} and \texttt{xillver} models account for the Fe K$\alpha$ emission at 6.4 keV. The residuals at $\sim 5.8$ indicate that additional emission is required red-ward of the line, suggesting line broadening. If broadening is included in the fit (\texttt{relxill, relxill+xillver}), the feature disappears.

The absorption in the \texttt{reflionx} fit at 7.1 keV disappears too when relativistic blurring is included. The last model (\texttt{relxill+xillver}) is included for completeness. It models reflection from both the inner (with relativistic broadening) and outer regions. Although, the additional component improves the $\chi^2$ slightly compared to \texttt{relxill} alone, its requirement is not very significant. The conclusion from Fig. \ref{fig:pg1211_sig} is that absorption features depend on the chosen continuum, and when the continuum is pinned down with \nustar coverage at high energies and the inclusion of relativistic reflection, no significant absorption is seen.

The analysis summarized in Fig. \ref{fig:pg1211_sig} was done using a combination of all the spectra (FPMA and FPMB from four observations) to obtain the best signal. We tested for any possible flux dependency by grouping observations 1 and 4 together and 2 and 3 together (see light curve in Fig. \ref{fig:lc_spec}-left), and the results do not change; no additional absorption lines are required by the data.

\subsection{Comparison with XMM-Newton}
The first {\it XMM-Newton} observation in 2001 showed an absorption line at $\sim 7.1$ keV in the observer's frame plus possible additional lines at lower energies \citep{2003MNRAS.345..705P}. We reanalyzed this {\it XMM-Newton} dataset using the same method discussed in section \ref{sec:line_search} and find the line at  $\sim7.5$ keV to be significant at the $\sim99\%$ level with an equivalent width of $104\pm52$ eV. The flux levels of the first {\it XMM-Newton} observation and the \nustar observation discussed here are comparable; therefore, we also assess whether an absorption line similar to that seen in {\it XMM-Newton} would have been detectable in the current \nustar data. We fitted the {\it XMM-Newton} PN spectrum with a model consisting of a power-law, relativistic reflection and an absorption line. We then used this model to fake \nustar data, and followed the procedure of section \ref{sec:line_search} to search for absorption line. The line is detected significantly in this simulated data showing that an absorption line would have been easily ($>8\sigma$ significance) detected if it was present in the data. Adding a Gaussian line to our best fit model (\texttt{relxill}) with the energy and the width of absorption line seen in the {\it XMM-Newton} spectrum provides a fit improvement of $\Delta\chi^2=3$ for one degree of freedom. The equivalent width of this line is $22\pm22$ eV.

Subsequent to the first observation, {\it XMM-Newton} observed \pg several times, and when we analyze these observations self-consistently, none of them show any significant absorption features when using a baseline model consisting of a power-law and a relativistic line.

\subsection {Relativistic Reflection Parameters}
\begin{figure}
\centering
 \includegraphics[width=190pt,clip ]{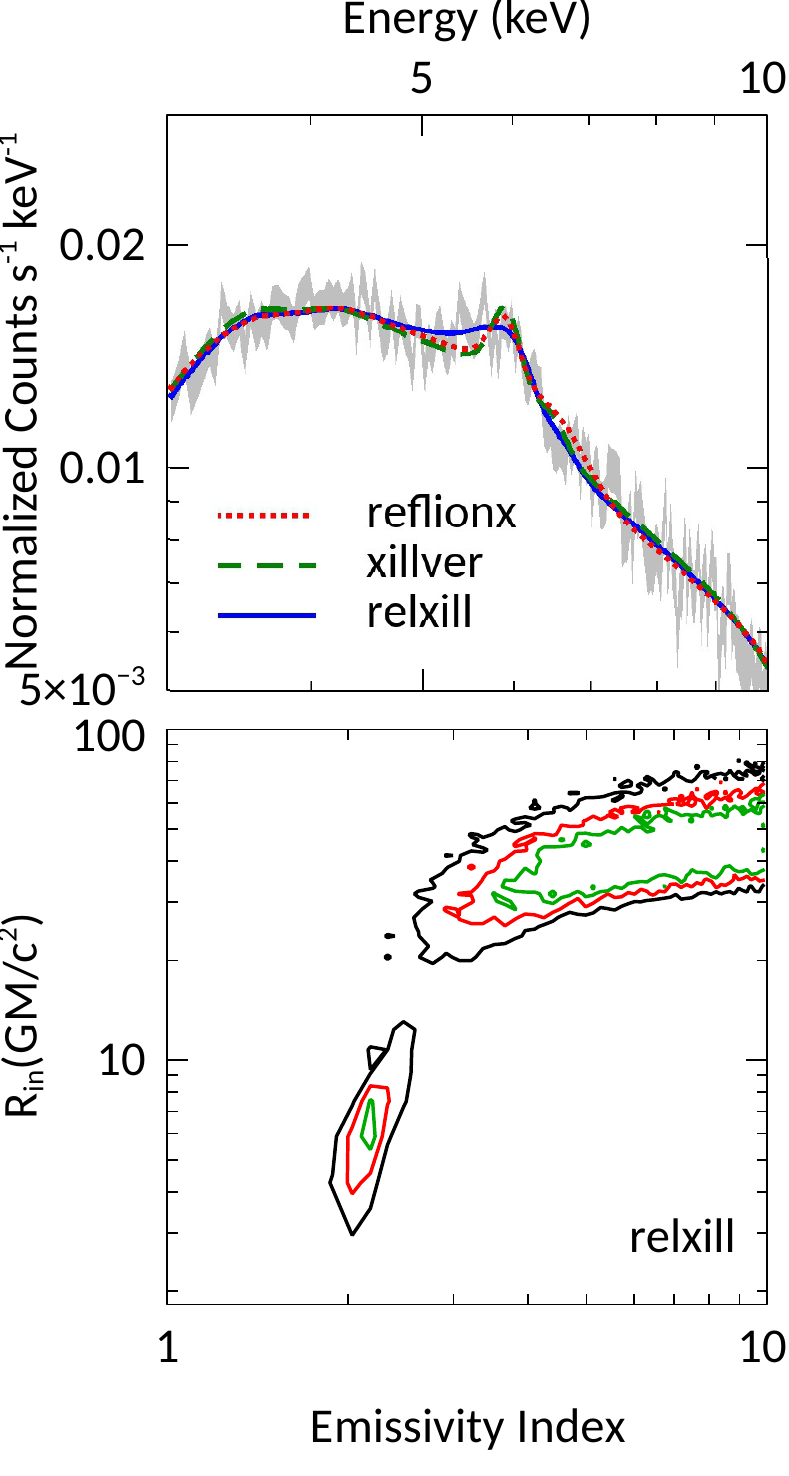} 
\caption{{\it Top:} The spectrum of \pg along with the best fit baseline models discussed in section \ref{sec:line_search} plotted in the 3--10 keV only where the models differ. The x-axis represents the rest-frame energy of the source. {\it Bottom:} Marginal probability density contours in the \texttt{relxill} model for the inner radius of emission and the emissivity index. The contours are for the 68, 90 and 99$\%$ confidence levels.}

\label{fig:pg1211_spec}
\end{figure}

The best fit model to the \nustar data, shown in Fig. \ref{fig:pg1211_spec}, consists of an absorbed power-law and relativistic reflection, both modeled with \texttt{relxill}. Only data below 10 keV are shown to highlight the differences among models. As was clear from Fig. \ref{fig:pg1211_sig}, the data show an excess at $\sim 5$ keV when fitted with simple reflection models. The addition of relativistic blurring (\texttt{relxill}) accounts for the excess. The data, however, do not allow very strong constraints on the spin of the black hole. The parameters of the best fit model are shown in Table \ref{tab:params}.

If we fit for the inner radius of emission instead of the black hole spin, we obtain the marginal probability contours shown in the bottom panel of Fig. \ref{fig:pg1211_spec}. This figure is obtained by running a Monte Carlo Markov Chain \footnote{We use xspec\_emcee available from: https://github.com/jeremysanders/xspec\_emcee } on the best fit \texttt{relxill} to obtain posterior probability distributions of the fit parameters and then marginalize over the parameters that are not of interest. The inner radius of emission and emissivity index are degenerate in the fit, but it is clear that although the extra broadening of the Fe K$\alpha$ line is required, it is not very large. The inner radius of emission ranges from $\sim 4$ to 40 gravitational radii ($r_g=GM/c^2$) depending on the disk emissivity. The solution with a steep emissivity and large radii might be unphysical, and although this data do not allow us to firmly locate the X-ray emission, the detection of reverberation time delays in this objects in the {\it XMM-Newton} data \citep{2011MNRAS.417L..98D} would favor the solution with a small radius for the emission.

\section {Discussion and Conclusions}
We have analyzed a deep 300~ks observation of the quasar \pg, obtained with \nustar in four time intervals spanning six months.  When fit
with a simple power-law model, the total spectrum shows clear
signatures of disk reflection (Fig. \ref{fig:lc_spec}-right).  Fits with
the newest, most sophisticated reflection models currently available
provide significantly improved fits.  The residuals to fits with simple reflection show an excess around $\sim 5$ keV indicating that relativistic blurring is required, clearly signaling that X-ray emission probes the very
innermost portion of the accretion flow in \pg.  The data do
not permit a strong constraint on the spin of the black hole, but when reverberation delays seen previously in {\it XMM-Newton} data are considered, a solution with a small inner radius ($\sim 4 r_g$, corresponding to a black hole spin of $\sim 0.5$) is supported by the data.  We thoroughly examined the \nustar data both in segments and in
total, and the spectra show no compelling evidence of a steady or variable
ultra-fast outflow seen previously in this source.  The limits are particularly tight once disk reflection has been modeled.

\nustar clearly has the sensitivity and spectral resolution
required to detect extraordinary outflows, when they are present.
Observations of PDS 456, for instance, reveal a likely P-Cygni profile
in that quasar, signaling a massive and powerful outflow from that
black hole into its host galaxy (Nardini et al. 2014, submitted).  \nustar
spectra from observations of NGC 1365 were able to disentangle
variable absorption, outflows, and disk reflection in that source,
leading to a robust measure of the black hole spin \citep{2013Natur.494..449R,2014ApJ...788...76W}.  At the opposite end of the mass scale, \nustar observations of the black hole candidate 4U~1630$-$472 revealed strong absorption that may be a UFO-like outflow, in addition to disk
reflection \citep{2014ApJ...784L...2K}. The true extent of these ultra-fast outflows and their persistence is unclear.

The non-detection of ultra-fast outflows in the prototypical object \pg in the current observation (and also previous observations subsequent to the first {\it XMM-Newton} detection) raises several questions on how common these might be in this and other sources. Systematic searches suggest that $\sim 40\%$ of objects observed with {\it XMM-Newton} and {\it Suzaku} show them \citep{2010A&A...521A..57T,2013MNRAS.430...60G}. This number could be much lower if a careful statistical search is performed that takes into account physical baseline models along with their observational uncertainties. This comes in the form of including the best fit uncertainties in simulating spectra during the Monte Carlo tests and also allowing for that uncertainty when fitting for narrow features at arbitrary energies in the spectrum. 

The true extent of UFO's could be affected by the significance of any single detection, which we have discussed, and also by their variability. A non-detection could simply be due to the fact they are transient in nature as suggested by the the simulations \citep{2004ApJ...616..688P}. The case of \pg shows that the UFO is seen in one out of eleven observations ({\it XMM-Newton}, {\it Suzaku}, {\it Chandra} and \nustar) between 2001 and 2014. With the caveat that we still have small numbers, if one assumes the observations are done randomly as far as the AGN is concerned, it appears that a variability argument would also suggest that the contribution of UFO's is an order of magnitude lower than previously inferred.

\section*{Acknowledgment}
This work made use of data from the {\em NuSTAR} mission, a project led by the California Institute of Technology, managed by the Jet Propulsion Laboratory, and funded by the National Aeronautics and Space Administration.

\bibliographystyle{astron}

\end{document}